\documentclass{article}
\usepackage{graphicx,amssymb}

\author{E.O. Babichev\footnote{babichev@inr.npd.ac.ru},
V.I. Dokuchaev\footnote{dokuchaev@inr.npd.ac.ru} ~and Yu.N.
Eroshenko\footnote{erosh@inr.npd.ac.ru}}
\title{The Accretion of Dark Energy onto a Black Hole}

\begin{document}
\maketitle

\begin{center}
{\em Institute For Nuclear Research of the Russian Academy of
Sciences \\
60th October Anniversary Prospect 7a, 117312 Moscow, Russia}
\end{center}

\begin{abstract}
The stationary, spherically symmetric accretion of dark energy
onto a Schwarzschild black hole is considered in terms of
relativistic hydrodynamics. The approximation of an ideal fluid is
used to model the dark energy. General expressions are derived for
the accretion rate of an ideal fluid with an arbitrary equation of
state $p=p(\rho)$ onto a black hole. The black hole mass was found
to decrease for the accretion of phantom energy. The accretion
process is studied in detail for two dark energy models that admit
an analytical solution: a model with a linear equation of state,
$p=\alpha(\rho-\rho_0)$, and a Chaplygin gas. For one of the
special cases of a linear equation of state, an analytical
expression is derived for the accretion rate of dark energy onto a
moving and rotating black hole. The masses of all black holes are
shown to approach zero in cosmological models with phantom energy
in which the Big Rip scenario is realized.
\end{abstract}

\section{INTRODUCTION}

In recent years, strong observational evidence that the Universe
is currently expanding with acceleration has been obtained. In the
Einstein theory of gravitation, this positive acceleration is
explained by the dominance of dark energy with a negative pressure
in the Universe [1-4]. Several theoretical models of dark energy
have been suggested: the vacuum energy (the cosmological constant
$\Lambda$) or such dynamical components as quintessence [5-10] and
k-essence [11-13]. Models with dynamical dark energy seem more
realistic, since tracker [14, 15], or attractor, solutions are
realized in them. Thus, the problem of fine tuning the parameters
of the Universe is solved [11-13].

A peculiar property of cosmological models with dark energy is the
possibility of a Big Rip [16, 17]: an infinite increase in the
scale factor of the Universe in a finite time. The Big Rip
scenario is realized in the case of dark energy, the so-called
phantom energy (for which $\rho+p<0$). In the Big Rip scenario,
the cosmological phantom energy density tends to infinity, and all
of the bound objects are torn apart up to subnuclear scales. It
should be noted, however, that the condition $\rho+p<0$ alone is
not enough for the Big Rip scenario to be realized [18]. Alam et
al. [19] analyzed data on distant supernovas in a
model-independent way and showed that the presence of phantom
energy with $-1.2<w<-1$ in the Universe at present is highly
likely. The quantum properties of the phantom energy in curved
spacetime were considered in [20]. The entropy of the Universe
filled with phantom energy was discussed in [21]. Models with
phantom energy are also used to construct mole burrows [22, 23].
The accretion of a scalar field onto a black hole from special
poten-tials $V(\phi)$ was considered in [24-29]. We use a
different approach to describe the accretion of dark energy onto a
black hole; more specifically, we model the dark energy by an
ideal fluid with a negative pressure.

In our recent paper [30] (see also [31]), we showed that the
masses of all black holes in the Universe with phantom energy
gradually decrease, and the black holes disappear completely by
the Big Rip. In this paper, we consider in detail the stationary
spherical accretion of dynamical dark energy onto a black hole.
The dark energy is modeled by an ideal fluid with a negative
pressure. The history of research on the accre-tion of an ideal
fluid onto a compact object begins with Bondi's classic paper
[32]. A relativistic generalization was made by Michel [33] (see
also [34-41] for further generalizations and supplements to
Michel's solution). Carr and Hawking [42] considered the accretion
of dust and radiation onto a black hole by solving the complete
system of Einstein equations and taking into account the back
reaction of the surrounding matter (see also [43] for a
description of the progress made in this area and for a discussion
of fundamental questions). Below, we obtain a solution for the
stationary accretion of a test relativistic ideal fluid with an
arbitrary equation of state $p(\rho)$ onto a Schwarzschild black
hole. Using this solu-tion, we show that the black hole mass
decreases during the accretion of phantom energy. The masses of
black holes can decrease during accretion in the case of phan-tom
energy due to the violation of the energy domi-nance condition
($\rho+p\geq0$) that underlies the theorem on the nondecreasing
area of the event horizon of a classical black hole [44].

This paper is structured as follows. In Section~2, we derive
general equations for the spherical accretion of an ideal fluid
and describe basic parameters of the steady energy flux onto a
black hole. We consider an arbitrary equation of state,
$w=p/\rho$, where the pressure $p$ can be positive (for ordinary
matter) and negative (for dark energy, including phantom energy
$w<-1$). Note that the parameter $w$ of the equation of state need
not be constant in our approach. Accretion causes the black hole
mass to change: the mass increases for $\rho+p>0$ and decreases
for $\rho+p<0$. The energy flux turns out to be completely
determined by the black hole mass $M$, the dark energy density at
infinity $\rho_\infty$, and the equation of state $p=p(\rho)$ only
if $0<\partial p/\partial\rho<1$. In this case, there is a
critical point that fixes the flux just as for an ordi-nary fluid.
When the condition $0<\partial p/\partial\rho<1$ is violated, the
dark energy flux onto a black hole can formally be arbitrary. For
$0<\partial p/\partial\rho<1$, we describe the method of
calculating the fluid parameters at the critical point and the
energy flux onto a black hole for given $M$, $\rho_{\infty}$, and
$p=p(\rho)$. In Section~3, we consider specific models of the
equation of state for dark energy. In the first model, we use a
simple equation of state with a linear density dependence of the
pressure. We consider the special cases of accretion of several
types of ideal fluid: ther-mal radiation, matter with an ultrahard
equation of state, dark energy with $\partial
p/\partial\rho\lessgtr 0$, and linear phantom energy. The
accretion rate of dark energy onto a moving black hole was
calculated for the special case of $\partial p/\partial\rho=1$. As
the second model, we investigate the accretion of a Chaplygin gas
onto a black hole. The evolution of the black hole mass in the
Universe with the Big Rip is considered in Section~4. The
possibility that the presence of phantom energy will lead the
Universe to the Big Rip in the future has been discussed in recent
years. The problem of the fate of black holes in this Universe is
solved in a rather unexpected way: black holes are not torn apart,
but disappear by the Big Rip due to the accretion of phantom
energy, irrespective of their initial masses. In Section~5, we
discuss the cor-respondence between the accretion of dark energy
modeled by an ideal fluid onto a black hole and the accretion of a
scalar field. The results obtained are briefly discussed in
Section~6.

\section{GENERAL EQUATIONS FOR SPHERICAL ACCRETION}
\label{Gen}

Let us consider the stationary, spherically symmetric accretion of
an ideal fluid that models the dark energy in the special case of
a negative pressure onto a black hole. The dark energy density is
assumed to be low enough for the metric to be a Schwarzschild one
with a high accuracy:
\begin{equation}
\label{metric} ds^2=\left(1-\frac{2M}{r}\right)dt^2 -
\left(1-\frac{2M}{r}\right)^{-1}dr^2 - r^2 (d\theta^2+\sin^2\theta
d\phi^2),
\end{equation}
where $M$ is the mass of the black hole, $r$ is the radial
coordinate, and $\theta$ and $\phi$ are the angular spherical
coordinates. We model the dark energy by an ideal fluid with the
energy-momentum tensor
\begin{equation}
\label{emt} T_{\mu\nu}=(\rho+p)u_\mu u_\nu - pg_{\mu\nu},
\end{equation}
where $\rho$ is the density, $p$ is the dark energy pressure, and
$u^\mu=dx^\mu/ds$ is the radial 4-velocity component. The pressure
is assumed to be an arbitrary function of the density,
$p=p(\rho)$. Integrating the zeroth (time) compo-nent of the
conservation law $T^{\mu\nu}_{\;\;\; ;\nu}=0$ yields the first
integral of motion for stationary, spherically symmetric accretion
(Bernoulli's relativistic equation or the energy equation):
\begin{equation}
 \label{eq1}
  (\rho+p)\left(1-\frac{2}{x}+u^2\right)^{1/2}x^2 u =C_1,
\end{equation}
where $x=r/M$, $u=dr/ds$, and $C_1$ is the constant determined
below. To find the second integral of motion, we use the equation
for the component of the energy-momentum tensor conservation law
along the 4-velocity
\begin{equation}
u_{\mu}T^{\mu\nu}_{\quad ;\nu}=0.
\end{equation}
In our case, this quation is [45]
\begin{equation}
\label{eq2} u^{\mu}\rho_{,\mu}+(\rho+p)u^{\mu}_{;\mu}=0.
\end{equation}
For the given equation of state
\begin{equation}
\label{p} p=p(\rho),
\end{equation}
the auxiliary function $n$ can be defined by the relation
\begin{equation}
 \label{n}
 \frac{d\rho}{\rho+p}=\frac{d n}{n}.
\end{equation}
The function $n$ is identical to the particle concentration for an
atomic gas, but it can also be used to describe a continuous
medium that does not consist of any particles. In this case, the
'concentration' $n$ is a formal auxiliary function. For an
arbitrary equation of state $p=p(\rho)$, we obtain a solution for
$n$ from Eq.~(\ref{n}):
\begin{equation}
  \label{n1}
  \frac{n(\rho)}{n_\infty}=\exp\left(\,\,\int\limits_{\rho_{\infty}}^{\rho}\frac{d\rho'}{\rho'+p(\rho')}\right),
\end{equation}
Using (\ref{n1}), we find the sought second integral of motion
from Eq.~(\ref{eq2}):
\begin{equation}
  \label{flux}
  \frac{n(\rho)}{n_\infty}ux^2=-A,
\end{equation}
where $n_\infty$ (the dark energy 'concentration' at infinity) was
introduced for convenience. In the case of a fluid flow directed
toward the black hole, $u=(dr/ds)<0$, and therefore, the numerical
constant $A>0$. From (\ref{eq1}) and (\ref{flux}), we can easily
obtain
\begin{equation}
  \label{energy}
  \frac{\rho+p}{n}\left(1-\frac{2}{x}+u^2\right)^{1/2} =C_2,
\end{equation}
where
\begin{equation}
\label{C2} C_2=\frac{\rho_\infty+p(\rho_\infty)}{n(\rho_\infty)}.
\end{equation}

Let us now calculate the radial 4-velocity component and the fluid
density on the event horizon of the black hole, $r=2M$. Setting
$x=2$, we obtain from Eqs. (\ref{flux}), (\ref{energy}) and
(\ref{C2})
\begin{equation}
\label{hor} \frac{A}{4}\;\frac{\rho_{\rm H}+p(\rho_{\rm
H})}{\rho_\infty+p(\rho_\infty)}=\frac{n^2(\rho_{\rm
 H})}{n^2(\rho_{\infty})},
\end{equation}
where $\rho_{\rm H}$ is the density on the $x=2$ horizon. Thus,
having specified the density at infinity $\rho_\infty$, the
equation of state $p=p(\rho)$, and the flux $A$ and using
definition (\ref{n}) of the concentration, we can calculate the
fluid density $\rho_{\rm H}$ on the event horizon of the black
hole from (\ref{hor}). Given the density on the horizon $\rho_{\rm
H}$, we can easily determine the radial fluid 4-velocity component
on the horizon from (\ref{flux}):
\begin{equation}
\label{u_H} u_{\rm
H}=-\frac{A}{4}\;\frac{n(\rho_\infty)}{n(\rho_{\rm H})}.
\end{equation}

Below, we will see that the constant $A$, which defines the energy
flux onto the black hole, can be calculated for hydrodynamically
stable ideal fluids with $\partial p/\partial\rho>0$. This can be
done by determining the fluid parameters at the critical point.
Following Michel [33], we find the relationship between the
parameters at the critical point:
\begin{equation}
\label{cpoint} u_*^2=\frac{1}{2 x_*},\quad
V_*^2=\frac{u_*^2}{1-3u_*^2},
\end{equation}
where
\begin{equation}
\label{V} V^2=\frac{n}{\rho+p}\frac{d(\rho+p)}{dn} -1,
\end{equation}
Together with (\ref{n}), this yields
\begin{equation}
\label{V1} V^2=c_s^2(\rho),
\end{equation}
where $c_s^2=\partial p/\partial\rho$ is the square of the
effective speed of sound in the medium. We derive the following
relation for the critical point from Eqs. (\ref{cpoint}),
(\ref{V1}), (\ref{C2}), and (\ref{energy}):
\begin{equation}
 \label{rho_c}
 \frac{\rho_*+p(\rho_*)}{n(\rho_*)}=
 \left[1+3c_s^2(\rho_*)\right]^{1/2}
 \frac{\rho_\infty+p(\rho_\infty)}{n(\rho_\infty)},
\end{equation}
which fixes the fluid density at the critical point $\rho_*$ for
an arbitrary equation of state $p=p(\rho)$. Specifying $\rho_*$
and using (\ref{n1}), we can determine $n(\rho_*)$. Accordingly,
the quantities $x_*$ and $u_*$ can be calculated from
(\ref{cpoint}) and (\ref{V1}). As a result, the numerical constant
$A$ can be calculated by substituting the derived quantities into
(\ref{flux}). For $c_s^2<0$ or $c_s^2>1$, no critical point exists
beyond the event horizon of the black hole ($x_*>1$), implying
that the dark energy flux onto the black hole depends on the
initial conditions for an unstable ideal fluid $c_s^2<0$ or a
'superluminal' fluid $c_s^2>1$. This result has a simple physical
interpretation: the accreted fluid has a critical point if its
speed increases from subsonic to supersonic values as it
approaches the black hole. In contrast, for $c_s^2$ or $c_s^2>1$,
the critical point either does not exist or is formally within the
event horizon of the black hole. It should also be noted that
fluids with $c_s^2<0$ are hydrodynamically unstable (see [46, 47]
for a discussion).

Equation (\ref{energy}), together with (\ref{p}), (\ref{n1}), and
(\ref{flux}), defines the accretion rate onto a black hole. These
equations are valid for an ideal fluid with an arbitrary equation
of state $p=p(\rho)$, in particular, for a gas of massless
particles (thermalized radiation) and a gas of massive particles.
For a gas of massive particles, the system of equations
(\ref{flux}) and (\ref{energy}) reduces to a similar system of
equations found by Michel [33]. It should be noted, however, that
Eqs. (\ref{p}), (\ref{n1}), (\ref{flux}), and (\ref{energy}) are
also valid for dark energy, including phantom energy with
$\rho+p<0$. In these cases, the concentration $n(\rho)$ is
positive for any $\rho$, while the constant $C_2$ in Eq.
(\ref{energy}) is negative.

The rate of change in the black hole mass (the energy flux onto
the black hole) through accretion is
$$\dot M=-4\pi r^2T_0^{\;r}.$$
Using (9) and (10), this expression can be rewritten as [30]
\begin{equation}
 \label{evol}
 \dot{M}=4\pi A M^2 [\rho_{\infty}+p(\rho_{\infty})].
\end{equation}
It follows from Eq. (18) than the mass of the black hole increases
as it accretes the gas of particles when $p>0$, but decreases as
it accretes the phantom energy when $p+\rho<0$. In particular,
this implies that the black hole masses in the Universe filled
with phantom energy must decrease. This result is general in
nature. It does not depend on the specific form of the equation of
state $p=p(\rho)$; only the satisfaction of the condition
$p+\rho<0$ is important. The physical cause of the decrease in the
black hole mass is as follows: the phantom energy falls to the
black hole, but the energy flux associated with this fall is
directed away from the black hole. If we ignore the cosmological
evolution of the density $\rho_\infty$, then we find the law of
change in the black hole mass from (18) to be
\begin{equation}
 \label{m}
 M=M_i\left(1-\frac{t}{\tau}\right)^{-1},
\end{equation}
where $M_i$ is the initial mass of the black hole, and $\tau$ is
the evolution time scale:
\begin{equation}
 \label{tau} \tau=1/\left\{4\pi A M_i
 [\rho_\infty+p(\rho_\infty)]\right\}.
\end{equation}

\section{ANALYTICAL ACCRETION MODELS}

\subsection{Model of a Linear Equation of State}

Let us consider the model of dark energy with a linear density
dependence of pressure [30]:
\begin{equation}
 \label{p1}
 p=\alpha(\rho - \rho_0),
\end{equation}
where $\alpha$ and $\rho_0$ 0 are constants. Among the other
cases, this model describes an ultrarelativistic gas ($p=\rho/3$),
a gas with an ultrahard equation of state ($p=\rho$), and the
simplest model of dark energy ($\rho_0=0$ and $\alpha<0$). The
quantity $\alpha$ is related to the parameter $w=p/\rho$ of the
equation of state by $w=\alpha(\rho-\rho_0)/\rho$.

An equation of state with $w=const<0$ throughout the cosmological
evolution is commonly used to analyze cosmological models. The
matter with such an equation of state is hydrodynamically unstable
and can exist only for a short period. Our equation of state (21)
for $\alpha>0$ does not have this shortcoming. For $\alpha>0$, it
also allows the case of hydrodynamically stable phantom energy to
be described, which is not possible when using an equation of
state with $w=const<-1$. In the real Universe, the equation of
state changes with time (i.e., $w$ depends on $t$). Therefore,
Eq.~(21) has the meaning of an approximation to the true equation
of state only in a limited $\rho$ range. From the physical point
of view, the condition $\rho>0$ must be satisfied for any equation
of state in a comoving frame of reference. In particular, the
state of matter with $\rho=0$, but $p\neq0$, is physically
unacceptable. The corresponding constraints for the equation of
state (21) are specified by conditions (29) and (30) given below.

For $\alpha<0$, there is no critical point for the accreted fluid
flow. For $\alpha>0$, using (14) and (16), we obtain the
parameters of the critical point
\begin{equation}
\label{cpoint1} x_*=\frac{1+3\alpha}{2\alpha},\quad
u_*^2=\frac{\alpha}{1+3\alpha}.
\end{equation}
Note that the parameters of the critical point (22) in the linear
model (21) are determined only by $\partial p/\partial\rho=\alpha$
and do not depend on $\rho_0$, which fixes the physical nature of
the fluid under consideration: a relativistic gas, dark energy, or
phantom energy. Note also that no critical point exists beyond the
event horizon of the black hole for $\alpha>1$ (this corresponds
to a nonphysical situation with a superluminal speed of sound).

Let us calculate the constant $A$, which defines the energy flux
onto the black hole. We find from (8) that
\begin{equation}
  \label{n2}
  \frac{n}{n_\infty}=\left|\frac{\rho_{\rm
  eff}}{\rho_{\rm eff,\infty}}\right|^{1/(1+\alpha)},
\end{equation}
where we introduced the effective density
$$\rho_{\rm eff}\equiv\rho+p= -\rho_0 \alpha +(1+\alpha)\rho$$
Using (17), we obtain
\begin{equation}
\label{rho_eff_c} \left(\frac{\rho_{\rm eff_*}}{\rho_{\rm
eff,\infty}}\right)^{\alpha/(1+\alpha)}= (1+3\alpha)^{1/2},
\end{equation}
where $\rho_{\rm eff_*}$ and $\rho_{\rm eff,\infty}$, are the
effective densities at the critical point and at infinity,
respectively. Substituting (24) into (23) and using (9), we obtain
for the linear model
\begin{equation}
\label{A1}
A=\frac{(1+3\alpha)^{(1+3\alpha)/2\alpha}}{4\alpha^{3/2}}.
\end{equation}
It is easy to see that $A\ge 4$ for $0<\alpha<1$. $A=4$ for
$\alpha=1$ (this corresponds to $c_s=1$); i.e., the constant $A$
is on the order of 1 for relativistic speeds of sound. Using (25),
we obtain from (20)
\begin{equation}
\label{tau1} \tau=\left[\pi M_i(\rho_\infty+p_\infty)
\frac{(1+3\alpha)^{(1+3\alpha)/2\alpha}}{\alpha^{3/2}}\right]^{-1}.
\end{equation}
To determine the fluid density on the event horizon of the
black hole, we substitute (23) into (12) to yield
\begin{equation}
  \label{rho_eff_H}
  \rho_{\rm H}=\frac{\alpha\rho_0}{1+\alpha}+
  \left(\rho_\infty-\frac{\alpha\rho_0}{1+\alpha}\right)\left(\frac{A}{4}\right)^{(1+\alpha)/(1-\alpha)},
\end{equation}
where $A$ is given by (25). For $0<\alpha<1$, the effective
density on the horizon $\rho_{\rm eff,H}$ cannot be lower than
$\rho_{\rm eff,\infty}$. The radial 4-velocity component on the
horizon can be found from (13) and (27):
\begin{equation}
\label{u_H_1} u_{\rm
H}=-\left(\frac{A}{4}\right)^{-\alpha/(1-\alpha)},
\end{equation}
The value of $u_{\rm H}$ changes from $1$ to $1/2$ for
$0<\alpha<1$.

The linear model (21) describes the phantom energy when
$\rho_\infty/\rho_0<\alpha/(1+\alpha)$. In this case, $\rho+p<0$.
However, the requirement that the density $\rho$ be nonnegative
should be taken into account. This parameter can formally be
negative in the range $0<\alpha\leq 1$. Such a nonphysical
situation imposes a constraint on the linear model (21) under
consideration. For a physically proper description of the
accretion process, we must require that the density $\rho$ be
nonnegative. We obtain the following constraint on the validity
range of the linear model from (27) for hydrodynamically stable
phantom energy:
\begin{equation}
  \label{goodph}
  \frac{\alpha}{1+\alpha}\left[1-\left(\frac{A}{4}\right)^{-(1+\alpha)/(1-\alpha)}\right]<
  \frac{\rho_\infty}{\rho_0}<
  \frac{\alpha}{1+\alpha}.
\end{equation}
As follows from (29), at a given $\alpha$, we can always choose
the parameters $\rho_0$ and $\rho_\infty$ in such a way that
$\rho>0$ for any $r>2 M$.

On the other hand, model (21) describes the quintessence (not the
phantom energy) for the entire $r$ range only if $p<0$.
Consequently, a physically proper description of the quintessence
can be obtained from (27) if
\begin{equation}
  \label{goodde}
  \frac{\alpha}{1+\alpha}<
  \frac{\rho_\infty}{\rho_0}<
  \frac{\alpha}{1+\alpha}\left[\frac{1}{\alpha}+\left(\frac{A}{4}\right)^{-(1+\alpha)/(1-\alpha)}\right].
\end{equation}

For some of the specific choices of $\alpha$ (more specifically,
for $\alpha=1/3$, $1/2$, $2/3$, and $1$), $\rho(x)$ and $u(x)$ can
be calculated analytically (see the Appendix for details on these
calculations). In Figs.~1 and 2, the radial 4-velocity component,
$u$, and the density normalized to the density at infinity,
$\rho/\rho_{\infty}$ are plotted against the coordinate $x=r/2M$.
\begin{figure}[t]
\includegraphics[angle=-90,width=0.99\textwidth]{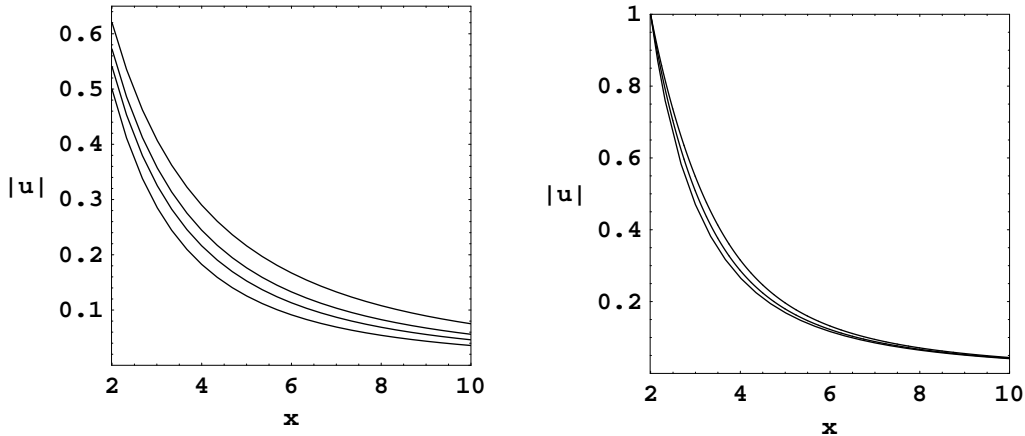}
  \caption{\label{Figul}
  Accreted fluid velocity $u$ in the linear model (21)
versus radial coordinate $x$.}
\end{figure}

\begin{figure}[t]
\includegraphics[angle=-90,width=0.99\textwidth]{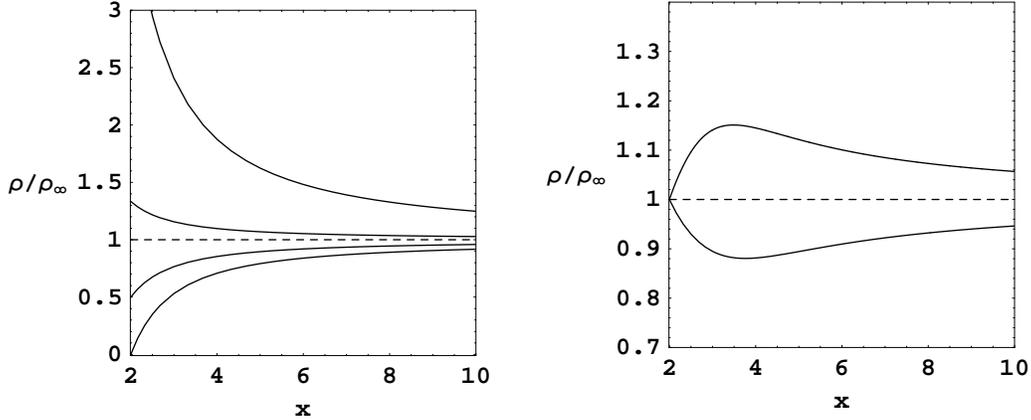}
  \caption{\label{Figrhol}
Accreted fluid density normalized to the density at infinity,
$\rho/\rho_{\infty}$, versus radial coordinate $x$ for the linear
model (21) (solid curves). The dashed lines indicate the density
of the $\Lambda$-term, $\rho/\rho_{\infty}$.
  }
\end{figure}

Figure 1 (left) shows the plots of $u(x)$ for hydrodynamically
stable fluids with $c_s^2>0$ at $\alpha=1/3$, $1/2$, $2/3$, and
$1$ (the curves are arranged from top to bottom, respectively).
Figure 1 (right) shows the plots of $u(r)$ for $c_s^2<0$ at
$\alpha=-1.1$, $-2$, and $-1/2$ (the curves are also arranged from
top to bottom). For this case, we chose the boundary condition
$u_{\rm H}=1$ on the horizon. Figure 2 (left) shows the plots of
$\rho/\rho_{\infty}$ for a hydrodynamically stable fluid with
$\alpha=1$ for various cases: $\rho_0=0$ (the model of neutron
star matter); $\rho_0/\rho_\infty=16/9$ (the linear model of
nonphantom dark energy); $\rho_0/\rho_\infty=7/3$ (the linear
model of phantom energy); and $\rho_0/\rho_\infty=7/3$ (the linear
model of phantom energy with $\rho_{\rm H}=0$) (the curves are
arranged from top to bottom, respectively). Figure 2 (right) shows
the plots of $(\rho/\rho_\infty)$ for various cases: $\alpha=-2$,
$\rho_0=0$, and $A=4$ (the linear model of phantom energy, the
upper curve); and $\alpha=-1/2$, $\rho_0=0$, and $A=4$ (the linear
model of nonphantom energy, the lower curve). For this case, we
chose the velocity $|u_{\rm H}|=1$ on the horizon.

\subsection{Accretion onto a Moving and Rotating Black Hole}

Let us consider the accretion onto a moving and rotating black
hole in the special case of a linear equation of state with
$\alpha=1$. The condition $\alpha=1$ allows an exact analytical
expression to be derived for the accre-tion rate of dark energy
onto a black hole.

For $\alpha=1$, we easily find from (23) that
\begin{equation}
  \label{n2a}
  \frac{n}{n_\infty}=\left|\frac{\rho_{\rm
  eff}}{\rho_{\rm eff,\infty}}\right|^{1/2}.
\end{equation}
We obtain the following continuity equation for the particle
concentration from (5):
$$
(n u^\mu)_{;\mu}=0.
$$
We can introduce the scalar field $\phi$ in terms of which the
fluid velocity can be expressed as follows (there is no torsion in
the fluid):
\begin{equation}
  \label{phi}
  \frac{\rho+p}{n}\;u_\mu=\phi_{,\mu}.
\end{equation}
We derive an equation for the auxiliary function $\phi$ by using
Eqs. (31) and (32),
\begin{equation}
  \label{phi1}
  \phi^\mu_{;\mu}=0.
\end{equation}
Exactly the same equation arises in the problem of the accretion
of a fluid with the equation of state $p=\rho$ [41]. Thus, we
reduced the problem of a black hole moving in dark energy with the
equation of state $p=\rho-\rho_0$ to the problem of a fluid with
an extremely hard equation of state, $p=\rho$. Using the method
suggested in [41], we obtain the mass evolution law for a moving
and rotating black hole immersed in dark energy with the equation
of state $p=\rho-\rho_0$:
\begin{equation}
 \label{evolmove}
 \dot{M}=4\pi(r_+^2+a^2) [\rho_{\infty}+p(\rho_{\infty})]
 u^0_{\rm BH},
\end{equation}
where
$$
r_+=M+(M^2-a^2)^{1/2}
$$
is the radius of the event horizon for a rotating black hole,
$a=J/M$ is the specific angular momentum of the black hole
(rotation parameter), and $u^0_\infty$ is the zeroth 4- velocity
component of the black hole relative to the fluid. Expression (34)
for $u^0_{\rm BH}=0$ reduces to (18) for a Schwarzschild ($a=0$)
black hole at rest.

\subsection{Chaplygin Gas}

Let us consider a Chaplygin gas with the following equation of
state as another example of the solvable model:
\begin{equation}
 \label{p3}
 p=-\frac{\alpha}{\rho},
\end{equation}
where $\alpha>0$. The range of parameters $\rho^2<\alpha$
represents the phantom energy with a superluminal speed of sound,
implying that the phantom energy flux onto the black hole is not
fixed by the condition of its passage through the critical point.
The case of $\rho^2>\alpha$ corresponds to dark energy with
$\rho+p>0$ and $0<c_s^2<1$. We can easily find from Eq. (8) that
\begin{equation}
  \label{n3}
  \frac{n}{n_\infty}=\left|\frac{\rho^2-\alpha}{\rho^2_\infty-\alpha}\right|^{1/2}.
\end{equation}
The density at the critical point can be calculated from (17) and
(36):
\begin{equation}
  \label{rho_3}
  \rho_*^2=4\rho_\infty^2 - 3\alpha.
\end{equation}
The velocity and the radial coordinate at the critical point are
given by
\begin{equation}
  \label{cpoint3}
  x_*=\frac{2\rho_\infty^2}{\alpha},\,
  u_*^2=\frac{\alpha}{4\rho_\infty^2}.
\end{equation}
We then find the constant $A$ from Eq. (9):
\begin{equation}
  \label{A3}
  A=4\left(\frac{\rho_\infty^2}{\alpha}\right)^{3/2}.
\end{equation}
For $0<c_s^2<1$, the constant $A$ cannot be smaller than $4$, as
in the case of the linear model. The evolution time scale of the
black hole mass without any cosmological change in the dark energy
density is given by
\begin{equation}
  \label{tau3} \tau=\left[8\pi M_i \frac{\rho_\infty^2}{\alpha}
  (\rho_\infty+p_\infty)\right].
\end{equation}
Note that Eqs. (36)-(39) are applicable only for dark energy with
$\rho+p>0$ and are invalid for phantom energy. On the black hole
horizon,
\begin{equation}
  \label{H_3}
  \rho_{\rm H}=\frac{A}{4}\rho_\infty,\, u_{\rm
  H}=-\frac{A}{4}\left[\frac{\rho_\infty^2-\alpha}{(A/4)^2\rho_\infty^2-\alpha}\right]^{1/2}.
\end{equation}
For $0<c_s^2<1$, the density on the horizon $\rho_{\rm H}$ cannot
be lower than $\rho_{\rm \infty}$, and $u_{\rm H}$ changes from
$1$ to $1/2$.
The Chaplygin gas density distribution can be
determined from the general equations (9) and (10):
\begin{equation}
  \label{rho_Ch}
   \rho=\rho_\infty\left(\frac{\rho_\infty^2}{\alpha}\right)^{3/2}
   \left(\frac{2}{x}-\frac{\alpha}{\rho_\infty^2}\right)^{-1/2}
   \left[\frac{16}{x^4}\left(1-\frac{\alpha}{\rho_\infty^2}\right)-
   \left(\frac{\alpha}{\rho_\infty^2}\right)^4\left(1-\frac{2}{x}\right)\right]^{1/2}.
\end{equation}
The velocity distribution $u(r)$ can be calculated by using Eqs.
(9), (36), and (42). In Figs.~3 and 4, the velocity $u$ and the
density normalized to the density at infinity,
$\rho/\rho_{\infty}$, are plotted against the coordinate $x=r/2M$.
\begin{figure}[t]
\includegraphics[angle=-90,width=0.99\textwidth]{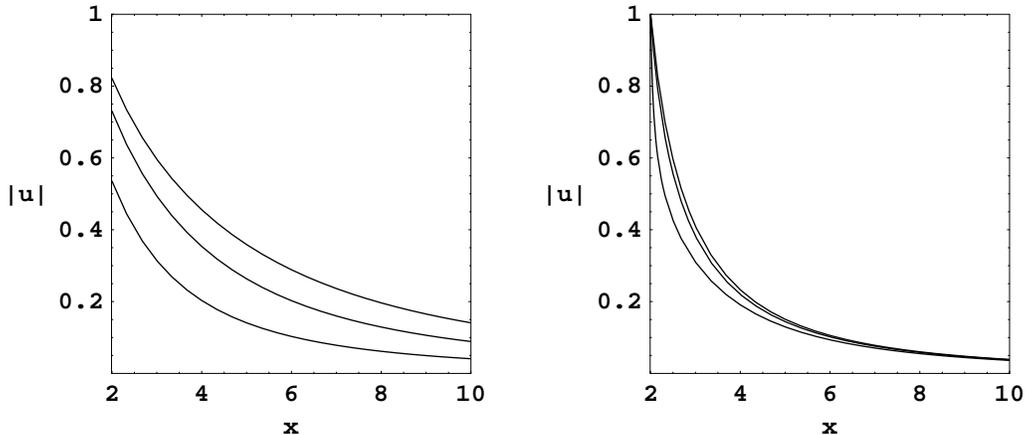}
  \caption{\label{FiguCh}
Velocity $u$ versus coordinate $x$ for Chaplygin gas [35]}
\end{figure}
\begin{figure}[t]
\includegraphics[angle=-90,width=0.99\textwidth]{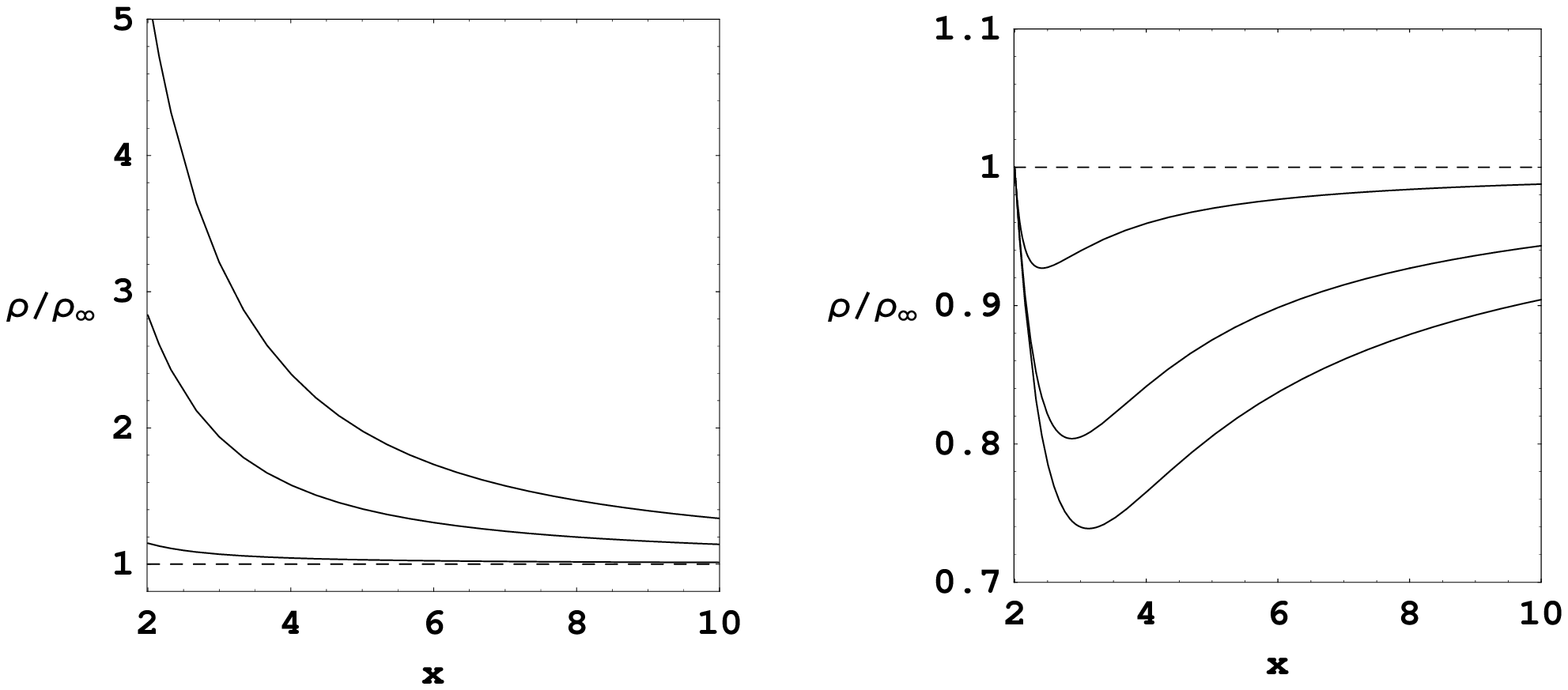}
  \caption{\label{FigrhoCh}
Density normalized to the density at infinity,
$\rho/\rho_{\infty}$, versus coordinate $x$ for model (35) (solid
curves). The dashed line indicates the normalized density of the
$\Lambda$-term, $\rho/\rho_\infty$.}
\end{figure}
Figure 3 (left) shows the plots of $u(r)$ for nonphantom dark
energy at $\rho_\infty^2/\alpha=3$, $2$, and $1.1$ (the curves are
arranged from top to bottom, respectively). Figure 3 (right) shows
the plots of $u(r)$ for phantom energy at
$\rho_\infty^2/\alpha=0.3$, $0.5$, and $0.9$ (the curves are also
arranged from top to bottom, respectively). In this case, the
boundary con-dition $u_{\rm H}=1$ is set on the horizon. Figure 4
(left) shows the plots of the normalized density,
$\rho/\rho_{\infty}$, for nonphantom dark energy at
$\rho_\infty^2/\alpha=3$, $2$, and $1.1$ (the curves are arranged
from top to bottom, respectively). Figure 4 (right) shows the
plots of the normalized density, $(\rho/\rho_\infty)$, for phantom
energy at $\rho_\infty^2/\alpha=0.9$, $0.5$, and $0.3$ (the curves
are also arranged from top to bottom, respectively). For this
case, we chose the boundary condition $u_{\rm H}=1$ on the
horizon.

\section{THE FATE OF BLACK HOLES DURING THE BIG RIP}

Let us now consider the evolution of black holes in the
cosmological Big Rip scenario, where the scale factor $a(t)$
increases to infinity in a finite time [16, 17]. For simplicity,
we take into account only the dark energy and disregard the other
forms of energy. In the linear model (21), the Big Rip takes place
at $\rho+p<0$ and $\alpha<-1$. The following relation can be
derived from Friedmann's equations in the case of a linear
equation of state:
$$
|\rho+p|\propto a^{-3(1+\alpha)}.
$$
Setting, for simplicity, $\rho_0=0$, we find the evolution law of
the phantom energy density in this Universe:
\begin{equation}
\label{sol3}
\rho_\infty=\rho_{\infty,i}\left(1-\frac{t}{\tau}\right)^{-2},
\end{equation}
where
\begin{equation}
\label{tau2}
\tau^{-1}=-\frac{3(1+\alpha)}{2}\left(\frac{8\pi}{3}\rho_{\infty,i}\right)^{1/2}
\end{equation}
Here, $\rho_{\infty,i}$ is the initial cosmological phantom
energy, and the initial time was chosen in such a way that the Big
Rip occurs at time $\tau$. We easily see from Eqs. (20) and (43)
that the Big Rip takes place at $\alpha\equiv\partial
p/\partial\rho<-1$. In general, the condition $\rho+p<0$ alone is
not enough for the cosmological evolution to be ended with the Big
Rip [18].

Using Eq. (43), we find the evolution of the black hole mass in
the cosmological Big Rip scenario from Eq.~(18):
\begin{equation}
 \label{mevol1}
 M=M_i\left(1+\frac{M_i}{\dot M_0 \;\tau}\;
 \frac{t}{\tau-t}\right)^{-1},
\end{equation}
where
\begin{equation}
\dot M_0=(3/2)\,A^{-1}|1+\alpha|,
\end{equation}

and $M_i$ is the initial mass of the black hole. At $\alpha=-2$
and a typical value of $A=4$ (which correspond to $u_{\rm H}=-1$),
$\dot M_0=3/8$. In the limit $t\to\tau$ (i.e., near the Big Rip),
the t dependence of the black hole mass becomes linear,
$M\simeq\dot M_0\,(\tau-t)$. When $t$ approaches $\tau$, the rate
of decrease in the black hole mass ceases to depend on the initial
black hole mass and the phantom energy density: In other words,
the masses of all black holes near the Big Rip are approximately
equal and approach zero. This implies that the accretion of
phantom energy dominates over the Hawking evaporation until the
black hole mass decreases to the Planck mass. Formally, however,
all black holes in the Universe completely evapo-rate during the
Hawking radiation in the Planck time before the Big Rip occurs.

\section{THE ACCRETION OF A SCALAR FIELD}
\label{Scalar}

In this section, we compare our calculations of the accretion of
an ideal fluid with similar calculations of the accretion of a
scalar (nonphantom) field onto a black hole [24-29]. The dark
energy is commonly modeled by a scalar field with a potential
$V(\phi)$. The approximation of an ideal fluid is rougher, since
the scalar field $\phi$ and $\partial_{\mu}\phi$ cannot be
unambiguously reproduced for given $\rho$ and $p$, which
characterize an ideal fluid. Despite this difference between the
scalar field and the ideal fluid, we will show that our results
are in close agreement with the corresponding calculations of the
accretion of a scalar field onto a black hole.

The Lagrangian of the scalar field is $L=K-V$, where $K$ is the
kinetic term and $V$ is the potential. For the standard choice of
the kinetic term
$$K=\phi_{;\mu}\phi^{;\mu}/2,$$
the corresponding energy flux onto the black hole is
$$
T_{0r}=\phi_{,t}\phi_{,r}.
$$
Jacobson [24] found a solution for the scalar field in the
Schwarzschild metric for a zero potential, $V=0$:
$$
\phi=\dot\phi_\infty[t+2M\ln(1-2M/r)],
$$
where $\phi_\infty$ is the scalar field at infinity. Frolov and
Kofman [26] showed that this solution is also valid for many
scalar fields with a nonzero potential V() under certain
conditions. For this solution
$$
T_0^{\;r}=-(2M)^2\dot\phi^2_\infty/r^2,
$$
and, accordingly,
$$
\dot M=4\pi(2M)^2{\dot{\phi}}^2_\infty.
$$

The energy-momentum tensor constructed using Jacobson's solution
is identical to the energy-momentum tensor for an ideal fluid with
an extremely hard equation of state, $p=\rho$, after the
substitution
$$
p_{\infty}\to\dot{\phi}_\infty^2/2,~~~~
\rho_{\infty}\to\dot{\phi}_\infty^2/2.
$$
This is not surprising, since the theory of a scalar field with a
zero potential, $V(\phi)$, is identical to the model of an ideal
fluid [48]. In view of this correspondence, we easily see
agreement between our result (18) for $\dot M$ in the case of
$p=\rho$ and the corresponding results from [24, 26].

The Lagrangian of the scalar field that describes the phantom
energy must have a negative kinetic term [16, 17], for example,
$$
K=-\phi_{;\mu}\phi^{;\mu}/2
$$
(see [49] for more general cases). In this case, the phantom
energy flux onto the black hole has the opposite sign,
$$
T_{0r}=-\phi_{,t}\phi_{,r},
$$
where $\phi$ is the solution of the same Klein-CGordon equation as
that for the standard scalar field, but with the substitution
$V\to-V$. For a zero potential, this solution is identical to
Jacobson's solution [24] obtained for a scalar field with a
positive kinetic term.

However, the Lagrangian with a negative kinetic term and
$V(\phi)=0$ does not describe the phantom energy. At the same
time, the solution for a scalar field with $V(\phi)=0$ is
identical to the solution for a positive constant potential,
$V_0=const$, which can be chosen in such a way that
$$
\rho=-\dot\phi^2/2+V_0>0.
$$
In this case, the scalar field describes the accreted phantom
energy with $\rho>0$ and $p<-\rho$, which leads to a decrease in
the black hole mass at the rate
$$
\dot M=-4\pi(2M)^2{\dot{\phi}}^2_\infty.
$$

A simple example of phantom cosmology (but without the Big Rip) is
realized by a scalar field with the potential
$$
V=m^2\phi^2/2,
$$
where $m\sim10^{-33}$~eV [50]. After a short transition period,
this cosmological model approaches an asymptotic state with
$$
H\simeq m\phi/3^{1/2},~~~\dot\phi\simeq2m/3^{1/2}.
$$
In the Klein-Gordon equation (with the substitution $V\to-V$
mentioned above), the term $m^2$ becomes equal to the other terms
only on the scale of the cosmological horizon, implying that, in
this case, Jacobson's solution is also valid. Calculations of the
corresponding energy flux onto the black hole yield
$$
\dot M=-4\pi(2M)^2\dot\phi^2_{\infty} =-64M^2m^2/3.
$$
For $M_0=M_{\odot}$ and $m=10^{-33}$~eV, the effective time of the
decrease in black hole mass is
$$
\tau=(3/64)M^{-1}m^{-2}\sim 10^{32}\mbox{~yr}.
$$

\section{DISCUSSION AND CONCLUSIONS}

In recent years, the concept of dark energy has been accepted and
extensively discussed in cosmology. The possible existence of dark
energy with a negative pressure leads to new cosmological
scenarios, including the exotic model of the Universe in which all
of the bound objects are destroyed and which dies itself as a
result of the Big Rip. To determine the fate of black holes in
this cosmological scenario, we considered the spherically
symmetric, stationary accretion of dark energy mod-eled by an
ideal fluid onto a black hole. We derived general equations for
the accretion of an ideal fluid with the equation of state
$p=p(\rho)$ onto a Schwarzschild black hole. In particular, these
equations can be used to describe the accretion of thermal
radiation, dark energy, and phantom energy. We also considered the
accretion onto a moving and rotating black hole in the special
case of an extremely hard equation of state, $p=\rho$. We
calculated the change in the black hole mass through accretion.
The black hole masses for $\rho+p>0$ were found to increase, as in
the usual case. However, a qualitatively new result was obtained
for phantom energy, i.e., for a medium with $\rho+p<0$. We found
that the black hole masses decrease in this situation. Using this
result, we solved the problem of the fate of black holes in a
universe that undergoes the Big Rip. It turns out that all black
holes in this Universe must decrease their masses and disappear
completely by the Big Rip. We also considered the correspondence
between the accretion of dark energy in the model of an ideal
fluid and the accretion of a scalar field.

\medskip

{\bf Acknowledgments:} This work was supported in part by the
Russian Foun-dation for Basic Research, project nos. 02-02-16762a,
03-02-16436a, and 04-02-16757a and by the Ministry of Science of
the Russian Federation, grants 1782.2003.2 and 2063.2003.2.

\section*{Appendix: Analytical Solutions for $\rho(x)$ and $u(x)$}

In the model under consideration (Section 3), analytical solutions
can be found for the dependence of the dark energy density and
accretion rate on radius $r$. Using Eq. (23) for the concentration
and Eq. (11) for the constant $C_2$, we derive the following
equation for $\rho_{\rm eff}$ from Eqs. (9) and (10)
\begin{equation}
\label{rho_an} {\left(\frac{\rho_{\rm eff}}{\rho_{\rm
eff,\infty}}\right)}^{2\alpha/(1+\alpha)}
\left[1-\frac{2}{x}+\frac{A^2}{x^4}{\left(\frac{\rho_{\rm
eff}}{\rho_{\rm eff,\infty}} \right)}^{-2/(1+\alpha)}\right]=1.
\end{equation}
Defining
\begin{equation}
\label{defy} y\equiv {\left(\frac{\rho_{\rm eff}} {\rho_{\rm
eff,\infty}}\right)}^{2/(1+\alpha)},
\end{equation}
we obtain the following equation from (47):
\begin{equation}
\label{y} y\left(1-\frac{2}{x}\right) - y^{1-\alpha} +
\frac{A^2}{x^4}=0,
\end{equation}
which can be solved analytically for certain values of $\alpha$.
For $\alpha=1/3$, Eq. (49) reduces to a cubic equation:
\begin{equation}
\label{z1} z^3\left(1-\frac{2}{x}\right)-z^2+\frac{A^2}{x^4}=0,
\end{equation}
where $z=y^{1/3}$. Solving this equation yields the fluid density
distribution for $\alpha=1/3$:
\begin{equation}
 \label{sol1}
 \rho=\frac{\rho_0}{4}+\left(\rho_{\infty}-\frac{\rho_0}{4}\right)
 \left[z+\frac{1}{3(1-2x^{-1})}\right]^2,
\end{equation}

where
\begin{equation}
  z=\left\{ \begin{array}{ll}
  2{\sqrt{\frac{a}{3}}}\,\cos\left(\frac{2\,\pi }{3}
  -\frac{\beta}{3}\right),& 2\leq x\leq3,\\
  2{\sqrt{\frac{a}{3}}}\,\cos\left(\frac{\beta}{3}\right),& x>3,
  \end{array} \right.
  \label{zz}
\end{equation}
\begin{equation}
 \beta=\arccos\left[\frac{b}{2\,(a/3)^{3/2}}\right]
 \label{beta}
\end{equation}
and
\begin{equation}
 a=\frac{1}{3{\left( 1-2/x\right) }^2},\;
 b=\frac{2}{27{\left(1-2/x\right) }^3}-
 \frac{108}{\left(1-2/x\right) x^4}.
\end{equation}
This solution corresponds to a thermalized photon gas in which the
photon mean free path is much smaller than the radius of the black
hole horizon, $\lambda_{fp}\ll2M$. In this situation, the photon
gas may be treated as an ideal fluid. In the opposite case,
$\lambda_{fp}\gg2M$, the photons are free particles, and their
accretion rate is determined by the well-known cross section for
the gravitational cap-ture of relativistic particles by a black
hole. The corresponding accretion rate is
$$
\dot M=27\pi M^2\rho_{\infty}
$$.

The case of $\alpha=2/3$ is similar to the case considered above.
We obtain the following equation instead of (50):
\begin{equation}
 z^3\left(1-\frac{2}{x}\right) - z + \frac{A^2}{x^4}=0,
 \label{z2}
\end{equation}

where again $z=y^{1/3}$. The fluid density distribution in this
case is
\begin{equation}
 \label{sol4}
 \rho=\frac{2}{5}\,\rho_0+\left(\rho_{\infty}-\frac{2}{5}\,\rho_0\right)
 z^{5/2},
\end{equation}
where $z$ is given by
\begin{equation}
  z=\left\{\begin{array}{ll}
  2{\sqrt{\frac{a}{3}}}\,\cos\left(\frac{2\,\pi }{3}
  -\frac{\beta}{3}\right),& 2\leq x\leq 9/4,\\
  2{\sqrt{\frac{a}{3}}}\,\cos\left(\frac{\beta}{3}\right),& x>9/4,
  \end{array} \right.
  \label{zz1}
\end{equation}
$\beta$ is defined by Eq. (53), and
$$
 a=\frac{1}{1-2/x},\quad
 b=-\frac{2187\,{\sqrt{3}}}{128\,\left( 1 - 2/x \right) \,x^4}.
$$
For $\alpha=1/2$, (49) is a quadratic equation and has a simple
analytical solution:
\begin{equation}
 \label{sol5}
 \rho=\frac{\rho_0}{3}+\left(\rho_{\infty}-\frac{\rho_0}{3}\right)
 z^{3/2},
\end{equation}
where
\begin{equation}
  z=\left\{\begin{array}{ll}
  \frac{1}{2}\left\{1-\left[1-3125\,\left(1-2/x\right)
  (16\,x^4)^{-1}\right]^{1/2}\right\}
  \left(1-2/x\right)^{-1},& 2\leq x\leq 5/2,\\
  \frac{1}{2}\left\{1+\left[1-3125\,\left(1-2/x\right)
  (16\,x^4)^{-1}\right]^{1/2}\right\}
  \left(1-2/x\right)^{-1},& x>5/2.
  \end{array} \right.
  \label{zz2}
\end{equation}
For $\alpha=1$, Eq. (49) is linear in $y$, which gives
\begin{equation}
 \label{sol2}
 \rho=\frac{\rho_0}{2}+\left(\rho_{\infty}-\frac{\rho_0}{2}\right)
 \left(1+\frac{2}{x}\right)\left(1+\frac{4}{x^2}\right).
\end{equation}


\begin{thebibliography}{99}

\bibitem{1} N. Bahcall, J. P. Ostriker, S. Perlmutter, and P. J.
Stein-hardt, Science 284, 1481 (1999).

\bibitem{2} A. Riess, A. V. Filippenko, P. Challis, et al., Astron. J. 116,
1009 (1998).

\bibitem{3} S. J. Perlmutter, G. Aldering, G. Goldhaber, et al.,
Astro-phys. J. 517, 565 (1999).

\bibitem{4} C. L. Bennett, M. Halpern, G. Hinshaw, et al., Astro-phys. J.,
Suppl. Ser. 148, 1 (2003).

\bibitem{5} C. Wetterich, Nucl. Phys. B 302, 668 (1988).

\bibitem{6} P. J. E. Peebles and B. Ratra, Astrophys. J. 325, L17 (1988).

\bibitem{7} B. Ratra and P. J. E. Peebles, Phys. Rev. D 37, 3406 (1988).

\bibitem{8} J. A. Frieman, C. T. Hill, A. Stebbins, and I. Waga, Phys. Rev.
Lett. 75, 2077 (1995).

\bibitem{9} R. R. Caldwell, R. Dave, and P. J. Steinhardt, Phys. Rev. Lett.
80, 1582 (1998).

\bibitem{10} A. Albrecht and C. Skordis, Phys. Rev. Lett. 84, 2076 (2000).

\bibitem{11} C. Armendariz-Picon, T. Damour, and V. Mukhanov, Phys. Lett. B
458, 209 (1999).

\bibitem{12} C. Armendariz-Picon, V. Mukhanov, and P. J. Steinhardt, Phys.
Rev. Lett. 85, 4438 (2000).

\bibitem{13} T. Chiba, T. Okabe, and M. Yamaguchi, Phys. Rev. D 62, 023511
(2000).

\bibitem{14} I. Zlatev, L. Wang, and P. Steinhardt, Phys. Rev. Lett. 82,
896 (1999).

\bibitem{15} P. Steinhardt, L. Wang, and I. Zlatev, Phys. Rev. D 59, 123504
(1999).

\bibitem{16} R. R. Caldwell, Phys. Lett. B 545, 23 (2002).

\bibitem{17} R. R. Caldwell, M. Kamionkowski, and N. N. Weinberg, Phys.
Rev. Lett. 91, 071301 (2003).

\bibitem{18} B. McInnes, J. High Energy Phys. 0208, 029 (2002); M.
Bouhmadi-Lopez and J. A. J. Madrid, astro-ph/ 0404540.

\bibitem{19} U. Alam, V. Sahni, T. D. Saini, and A. A. Starobinsky,
astro-ph/0311364.

\bibitem{20} S. Nojiri and S. D. Odintsov, Phys. Lett. B 562, 147 (2003).

\bibitem{21} I. Brevik, S. Nojiri, S. D. Odintsov, and L. Vanzo, hep-th/
0401073.

\bibitem{22} M. Visser, S. Kar, and N. Dadhich, Phys. Rev. Lett. 90, 201102
(2003).

\bibitem{23} P. F. Gonzalez-Diaz, Phys. Rev. D 68, 084016 (2003).

\bibitem{24} T. Jacobson, Phys. Rev. Lett. 83, 2699 (1999).

\bibitem{25} R. Bean and J. Magueijo, Phys. Rev. D 66, 063505 (2002).

\bibitem{26} A. Frolov and L. Kofman, J. Cosmol. Astrophys. Phys. 5, 9
(2003).

\bibitem{27} W. G. Unruh, Phys. Rev. D 14, 3251 (1976).

\bibitem{28} L. A. Urena-Lopez and A. R. Liddle, Phys. Rev. D 66, 083005
(2002).

\bibitem{29} M. Yu. Kuchiev and V. V. Flambaum, gr-qc/0312065.

\bibitem{30} E. O. Babichev, V. I. Dokuchaev, and Yu. N. Eroshenko, Phys.
Rev. Lett. 93, 021102 (2004).

\bibitem{31} E. O. Babichev, V. I. Dokuchaev, and Yu. N. Eroshenko, Class.
Quantum Grav. 22, 143 (2005).

\bibitem{32} H. Bondi, Mon. Not. R. Astron. Soc. 112, 195 (1952).

\bibitem{33} F. C. Michel, Astrophys. Space Sci. 15, 153 (1972).

\bibitem{34} B. J. Carr and S. W. Hawking, Mon. Not. R. Astron. Soc. 168,
399 (1974).

\bibitem{35} M. C. Begelman, Astron. Astrophys. 70, 583 (1978).

\bibitem{36} D. Ray, Astron. Astrophys. 82, 368 (1980).

\bibitem{37} K. S. Thorne, R. A. Flammang, and A. N. Zytkow, Mon. Not. R.
Astron. Soc. 194, 475 (1981).

\bibitem{38} E. Bettwieser and W. Glatzel, Astron. Astrophys. 94, 306
(1981).

\bibitem{39} K. M. Chang, Astron. Astrophys. 142, 212 (1985).

\bibitem{40} U. S. Pandey, Astrophys. Space Sci. 136, 195 (1987).

\bibitem{41} L. I. Petrich, S. L. Shapiro, and S. A. Teukolsky, Phys. Rev.
Lett. 60, 1781 (1988).

\bibitem{42} B. J. Carr and S. W. Hawking, Mon. Not. R. Astron. Soc. 168,
399 (1974).

\bibitem{43} E. Bettwieser and W. Glatzel, Astron. Astrophys. 94, 306
(1981).

\bibitem{44} S. W. Hawking and G. F. R. Ellis, The Large Scale Struc-ture
of Space-Time (Cambridge Univ. Press, Cambridge, 1973), Chap. 4.3.

\bibitem{45} C. W. Misner, K. S. Thorne, and J. A. Wheeler, Gravita-tion
(Freeman, San Francisco, 1973).

\bibitem{46} J. C. Fabris and J. Martin, Phys. Rev. D 55, 5205 (1997).

\bibitem{47} S. M. Carroll, M. Hoffman, and M. Trodden, Phys. Rev. D 68,
023509 (2003).

\bibitem{48} V. N. Lukash, Zh. Eksp. Teor. Fiz. 79, 1601 (1980) [Sov. Phys.
JETP 52, 807 (1980)].

\bibitem{49} P. F. Gonzalez-Diaz, Phys. Lett. B 586, 1 (2004).

\bibitem{50} M. Sami and A. Toporensky, Mod. Phys. Lett. A 19, 1509 (2004).


\end{thebibliography}
\end{document}